\documentstyle[11pt,twoside,jltp,psfig,epsf]{article}
\runninghead{Andersen et al.}{Out-of-plane instability and 
electron-phonon contribution to ...}

\begin{document}

\title{Out-of-plane instability and electron-phonon contribution to $s$- and $d$%
-wave pairing in high-temperature superconductors; LDA linear-response
calculation for doped CaCuO$_2$ and a generic tight-binding model.}
\author{O. K. Andersen, S. Y. Savrasov, O. Jepsen, and A.I.Liechtenstein%
\address{Max--Planck Institut
f\"ur Festk\"orperforschung, 70569 Stuttgart, Germany}}

\begin{abstract}
The equilibrium structure, energy bands, phonon dispersions, and $s$- and $d$%
-channel electron-phonon interactions (EPIs) are calculated for the
infinite-layer superconductor CaCuO$_2$ doped with 0.24 holes per CuO$_2$.
The LDA and the linear-response full-potential LMTO method were used. In the
equilibrium structure, oxygen is found to buckle slightly out of the plane
and, as a result, the characters of the energy bands near $\epsilon _F$ are
found to be similar to those of other optimally doped HTSCs. For the EPI we
find $\lambda _s\sim $0.4, in accord with previous LDA calculations for YBa$%
_2$Cu$_3$O$_7.$ This supports the common belief that the EPI mechanism alone
is insufficient to explain HTSC. $\lambda _{x^2-y^2}$ is found to be
positive and nearly as large as $\lambda _s.$ This is surprising and
indicates that the EPI could enhance some other $d$-wave pairing mechanism.
Like in YBa$_2$Cu$_3$O$_7,$ the buckling modes contribute significantly to
the EPI, although these contributions are proportional to the static
buckling and would vanish for flat planes. These numerical results can be
understood from a generic tight-binding model originally derived from the
LDA bands of YBa$_2$Cu$_3$O$_7$. In the future, the role of anharmonicity of
the buckling-modes and the influence of the spin-fluctuations should be
investigated.

PACS numbers: 74.72.Jt, 74.25.Kc, 63.20.Kr
\end{abstract}

\maketitle

%Include this space if you do not use sections in your document.
%\vspace{0.3in}

\section{INTRODUCTION}

The mechanism of high-temperature superconductivity in hole-doped CuO$_2$
materials remains a subject of vivid debate \cite{Plakida}. The Coulomb
repulsion between the electrons is poorly screened, at least for low doping
levels, and it may be responsible for the pairing \cite{corr}. This is
supported by experimental evidence that the symmetry of the paired state is $%
d_{x^2-y^2}$ with lobes in the direction of the CuO-bond. Nevertheless, the
results of Hubbard- and {\it t-J}-model calculations using realistic
parameters lead to a suspicion that something more than the Coulomb
repulsion is needed \cite{corr}. A large amount of experimental data such as
superconductivity-induced phonon renormalizations \cite{Cardona}, a large
isotope effect away from optimal doping \cite{Frank}, and phonon-related
features in the tunnelling spectra \cite{Vedeneev} show effects of the
electron-phonon interaction (EPI). It is therefore of interest to
investigate whether the EPI, which gives a pair-interaction, $V\left( {\bf %
k,k}^{\prime }\right) \propto -\left| g\left( {\bf k,k}^{\prime }\right)
\right| ^2,$ which is always negative (attractive), could support $%
d_{x^2-y^2}$-wave pairing with a gap-anisotropy like $\Delta \left( {\bf k}%
\right) \propto \cos \left( ak_x\right) -\cos \left( ak_y\right) $. For this
to occur, $\left| g\left( {\bf k,k}^{\prime }\right) \right| ^2$ must be
large when $\Delta \left( {\bf k}\right) $ and $\Delta \left( {\bf k}%
^{\prime }\right) $ have the same sign, and small when they have opposite
signs. Such an investigation is the main topic of the present paper.

We first report {\it ab initio }local density-functional (LDA) calculations
of the crystal, electronic, and phononic structure, and of the $s$- and $d$%
-channel EPIs for the infinite-layer compound CaCuO$_2$. Unfortunately, no
single-crystal measurements exist for an infinite-layer HTSC, and the
ceramic samples for which a $T_c$ of 110K was reported \cite{disc} may be
phase impure \cite{nosuper}. However, the infinite-layer structure is
uniquely simple for theoretical studies and high-temperature
superconductivity is generally believed to take place in CuO$_2$ planes. As
we shall see, also the LDA electronic structure of this material, properly
relaxed, is very similar to that of the stoichiometrically doped,
well-characterized HTSC, YBa$_2$Cu$_3$O$_7$. Our calculations are for 24 per
cent hole-doped CaCuO$_2$ with the hole charge being neutralized by a
homogeneous, negative background charge. We have chosen a doping of 0.24
holes per CuO$_2$ because this places the Fermi level at the uppermost
saddle-point in the LDA band structure, and because it agrees with the
experimental (average) doping level.

In the second part of the paper we present analytical low-energy bands,
originally derived from LDA calculations for YBa$_2$Cu$_3$O$_7$ \cite{An94},
and derive the EPI for the buckling-mode. This tight-binding model has
previously \cite{An95} been used to explain the in-plane dispersion of the
inter-plane hopping integral $t_{\perp }\left( {\bf k}_{\parallel }\right) $%
, which is a crucial ingredient of the inter-plane tunnelling model of HTSC 
\cite{PWA}.When augmented with a mean-field treatment of the Coulomb
repulsion $U$ between two holes in the same Cu $d_{x^2-y^2}$ orbital, the
model has correctly predicted the value of the inter-plane exchange coupling 
$J_{\perp }$ \cite{Monien}. In the present paper we consider the EPI in
optimally and overdoped HTSCs and shall assume that $U$ is screened
completely. We shall derive an analytical expression for the coupling $%
g\left( {\bf k,k+q}\right) $ of the electrons to the buckling mode for a
single plane and show that, for this mode, $\lambda $ for $d_{x^2-y^2}$-wave
pairing is positive definite. The analytical tight-binding model should also
be useful for further studies.

\section{LDA CALCULATIONS\ FOR CaCuO$_2$}

\subsection{Energy bands and equilibrium structure.}

In this section we discuss the electronic energy bands and the equilibrium
structure calculated with our full-potential linear-muffin-tin-orbital
(LMTO) method. The bands presented in Fig. 1 agree in detail with those of
previous calculations \cite{Novikov}, and they hardly change upon hole
doping by 10 per cent, except for a downwards shift of the Fermi energy to
the uppermost van Hove singularity (at R) as shown in the figure. The Fermi
surface (FS)\ for the doped compound (see Fig. 4a) has a square
(10)-oriented cross-section with the $\Gamma $XM-plane $\left( k_z\rm{=0}%
\right) $ and a (11)-oriented cross section with the ZRA-plane $\left( k_z%
\rm{=}\frac \pi c\right) $.

\begin{figure}
\centerline{\psfig{file=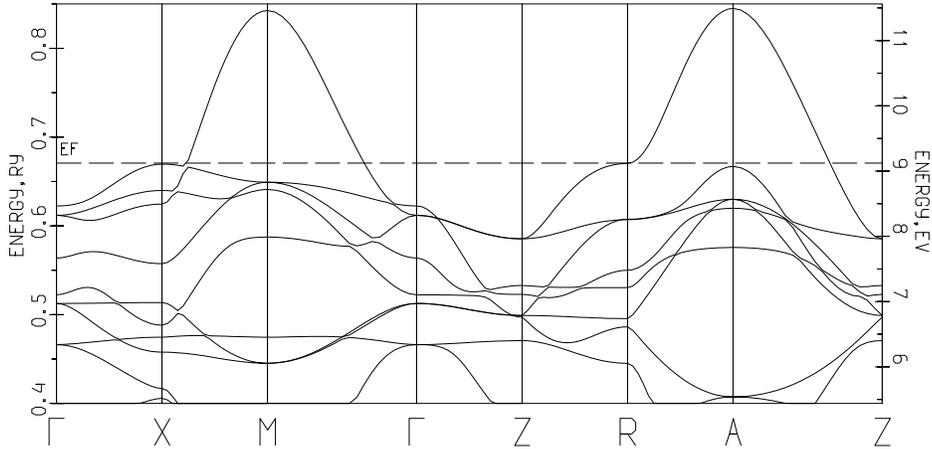,height=6cm}}
\caption{
Band structure of doped CaCuO$_2$ with
flat planes.
}
\label{fig:fig1}
\end{figure}

Although the band structure in Fig. 1 has antibonding in-plane $pd\sigma $%
-character (O$_x$-Cu$_{x^2-y^2}$-O$_y),$ like for all CuO$_2$-HTSCs, at and
above $\epsilon _F,$ there are strong {\em a}typical features just below $%
\epsilon _F:$ First of all, the uppermost band at X is an antibonding
out-of-plane $pd\pi $ band [the O$_z$-Cu$_{xz}$ band at $\left( \frac \pi
a,0,0\right) $ and the O$_z$-Cu$_{yz}$ band at $\left( 0,\frac \pi
a,0\right) ],$ whereas at R [$\left( \frac \pi a,0,\frac \pi c\right) $ and $%
\left( 0,\frac \pi a,\frac \pi c\right) ]$ the uppermost band is the usual $%
pd\sigma $-band. This is due to a relatively strong dispersion, $2t_{\perp
}^\pi \left( {\bf k}_{\parallel }\right) \cos ck_z,$ of the $pd\pi $ band in
the $z$-direction, a dispersion which is opposite to that, $-2t_{\perp
}^\sigma \left( {\bf k}_{\parallel }\right) \cos ck_z,$ of the $pd\sigma $
band. Secondly, near A the top of the antibonding in-plane $pd\pi $ band $%
\left( \rm{O}_y\rm{-Cu}_{xy}\rm{-O}_x\right) $nearly reaches $\epsilon
_F.$

Our total-energy calculations however predict that this infinite-layer
structure is unstable with respect to the ${\bf q}$=0 oxygen out-of-phase
buckling mode (B$_{2u}$). We have performed frozen-phonon calculations for
both undoped and doped compounds and with both experimental and theoretical
lattice parameters. The results are shown in Fig. 2. We see that a
double-well potential exists for such buckling motions, when the system is
doped. The minimum is at 5$^{\circ }$ if we use the experimental values: $a$%
=7.297 a.u. and $c/a$=0.829, and at 7$^{\circ }$ if we use our theoretically
determined values: $a$=7.112 a.u. and $c/a$=0.895. Note that YBa$_2$Cu$_3$O$%
_7$ has a 7$^{\circ }$ static in-phase dimpling of the Cu-O planes, both
experimentally and theoretically.

\begin{figure}[t]
\centerline{\epsfysize=10cm \epsffile{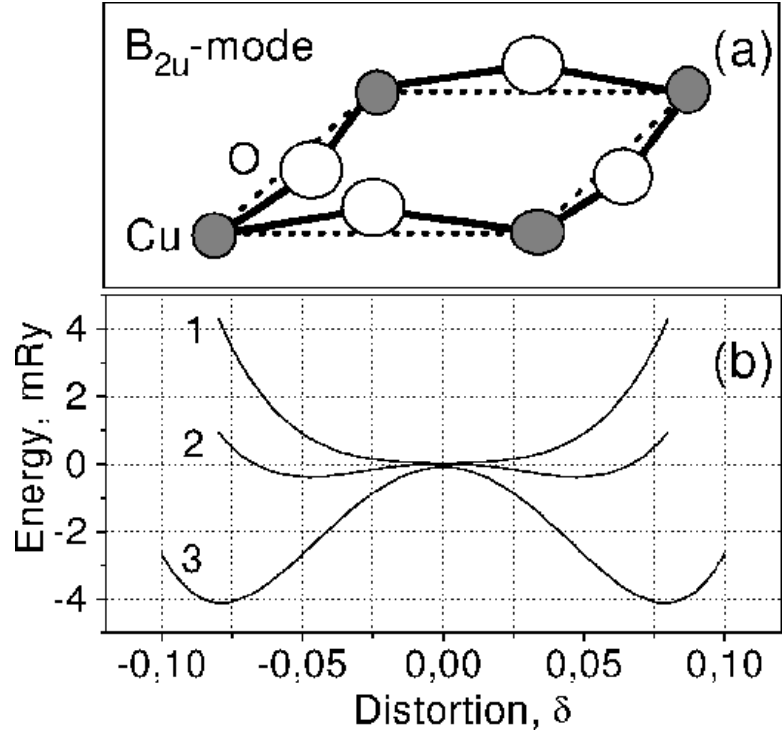}}
\caption{
(a) Unstable B$_{2u}$ optical mode
involving out--of--plane and out--of--phase displacements of oxygen atoms.
(b) Change in the total energy associated with the B$_{2u}$ mode. Curve 1
corresponds to the undoped compound calculated with the experimental lattice
constants $a_{exp}$ and $c_{exp}.$ Curve 2: Doped compound with $a_{exp}$, $%
c_{exp}.$ Curve 3: Doped compound with calculated lattice constants.
}
\label{fig:fig2}
\end{figure}

The buckling reduces the in-plane $t_{z,xz}$=$t_{z,yz}$ and $t_{y,xy}$=$%
t_{x,xy}$ hopping integrals and thereby causes the tops of the out-of-plane
and in-plane $pd\pi $ bands to move well below $\epsilon _F$. For the
theoretically stable structure, our bands in Fig. 3 and Fermi-surface cross
sections in Fig. 4b are similar to those of other cuprates. Comparing in
detail with the calculated FS of YBa$_2$Cu$_3$O$_7$ \cite{An94}, where the
dimple causes the saddlepoint at the Fermi level (that of the odd plane
band) to be bifurcated by $\Delta k$ to the positions $\left( \frac \pi a\pm
\Delta k,0,0\right) $, we see that the saddlepoint at the Fermi level in
CaCuO$_2$ is {\em not} bifurcated away from R $\left( \frac \pi a,0,\frac
\pi c\right) .$ The saddlepoints well below $\epsilon _F$ (that of the even
plane band in YBa$_2$Cu$_3$O$_7$ and that for $k_z$=0 in CaCuO$_2)$ {\em are}
however bifurcated away from X in both materials. (The difference between
the saddlepoints at $k_z$=$\frac \pi c$ and $k_z$=0 in CaCuO$_2$ is due to
the opposite $k_z$-dispersions of the $pd\sigma $ and $pd\pi $ bands.)

\begin{figure}
\centerline{\psfig{file=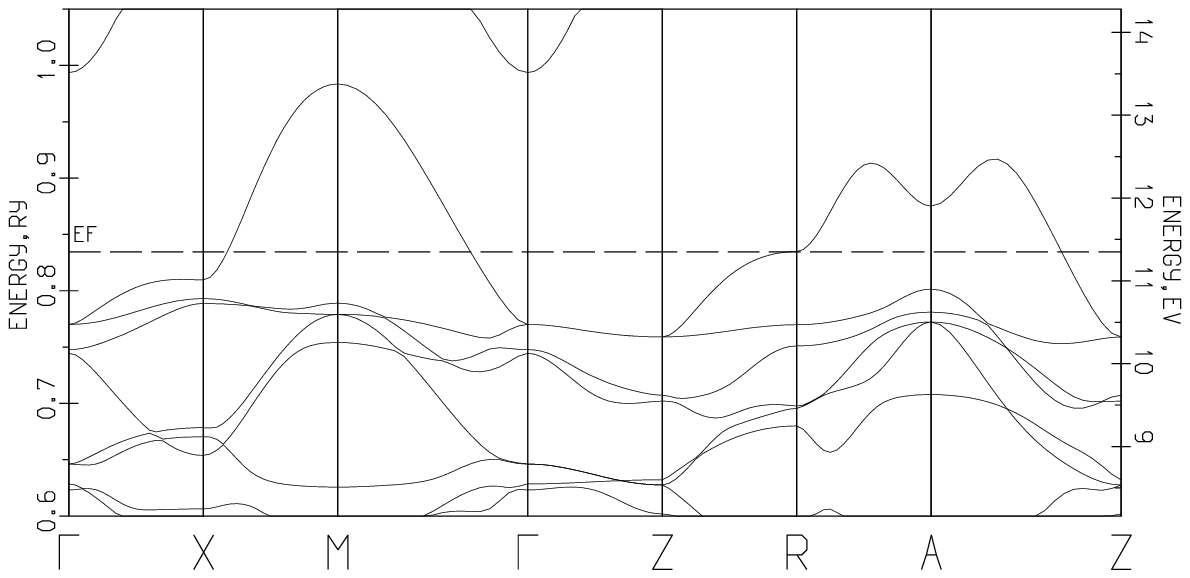,height=6cm}}
\caption{
Band structure of doped CaCuO$_2$ with
buckled planes.
}
\label{fig:fig3}
\end{figure}

By comparison of the FS cross-sections in Figs 4a and b, we see that the
static buckle squares up the (10)-oriented cross sections and thus creates 
{\em large, flat} (100)- and (010)-oriented pieces on the FS, centered near
respectively $\left( \frac{5\pi }{6a},0,0\right) $ and $\left( 0,\frac{5\pi 
}{6a},0\right) .$

\begin{figure}
\centerline{\psfig{file=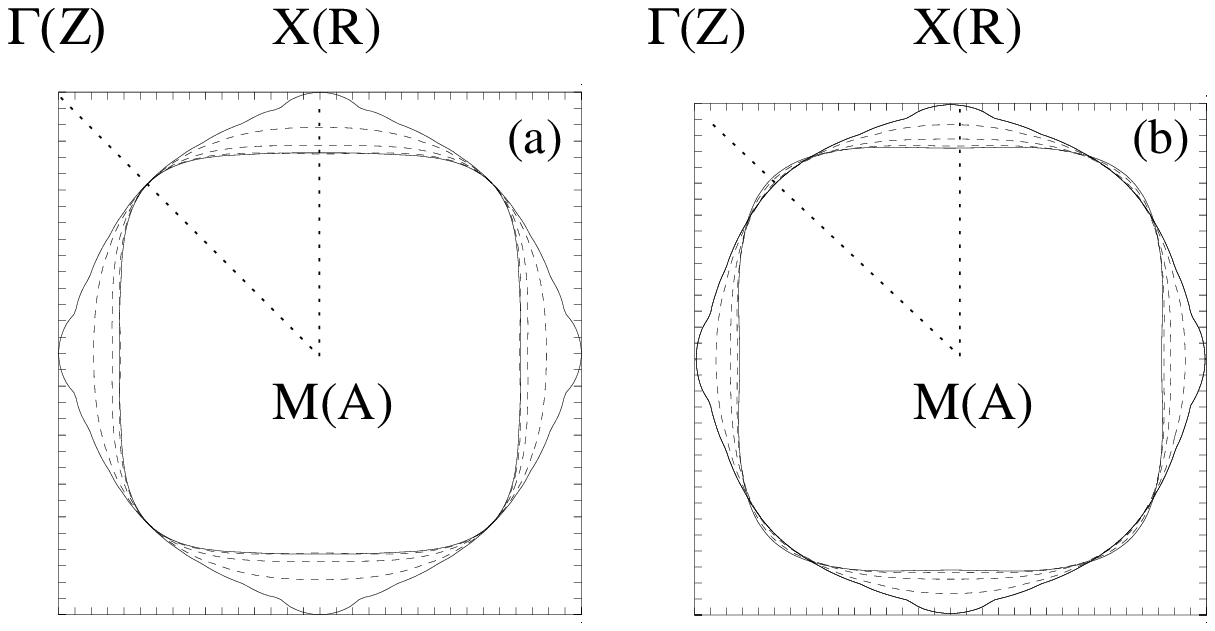,height=6cm}}
\caption{
Fermi surface cross sections of doped
CaCuO$_2$ with flat (a) and buckled (b) planes. The (10)--oriented cross
section is with the $\Gamma XM$ plane while (11)--oriented cross section is
with the $ZRA$ --plane. The dashed lines show intermediate cross sections.
}
\label{fig:fig4}
\end{figure}

The calculated instability suggests that out-of-plane buckling occurs
generally in such materials when the Fermi level is lowered to the top of
the $pd\pi $ bands. Inclusion of the Coulomb correlations, would presumably
cause such instabilities to occur at lower doping levels than we predict
with the LDA because, in a mean-field treatment, the energy of the
majority-spin Cu $d_{x^2-y^2}$ orbital is lowered. The striped
anti-ferromagnetic low-temperature-tetragonal (AF-LTT) phases recently found
by neutron scattering in $\frac 18$-doped La$_2$CuO$_4$ \cite{Tranquada} and
the local buckling instabilities found in 15 per cent doped La$_2$CuO$_4$ by
EXAFS \cite{EXAFS} are presumably examples of this. Only when pinned, do the
AF stripes seem to be decremental to superconductivity.

\subsection{Phonon spectrum}

We now present the phonon dispersions calculated for the doped compound. We
use newly developed linear-response full-potential LMTO method \cite{PHN}
whose accuracy was proven on lattice-dynamical, superconducting, and
transport properties for a large variety of metals \cite{EPI}. The
theoretical lattice constants were used in this calculation since the
shallowness of the B$_{2u}$-well calculated with the experimental
parameters casts serious doubts on the validity of the harmonic
approximation. The technical details of this calculation can be found
elsewhere \cite{EPIPRL}.

\begin{figure}[t]
\centerline{\epsfysize=10cm \epsffile{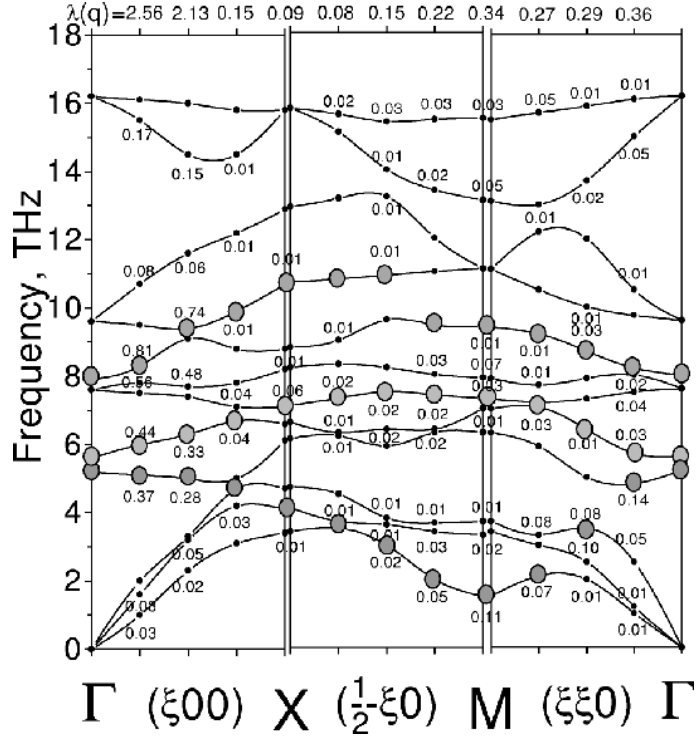}}
\caption{
Calculated phonon spectrum of doped
CaCuO$_2$ for $q_z$=0. Of the branches labelled by solid circles, the lowest
is a quadrupolar mode involving in-plane rotation of oxygen squares around
Cu and $z$-movement of Ca. The two uppermost branches labelled by solid
circles involve out-of-plane oxygen buckling. The numbers give the $s$%
-channel mode EPIs, $\lambda _{s,\nu }\left( q\right) ,$ and, at the top of
the figure, the sum over all modes, $\lambda _s\left( q\right) =\sum_\nu
\lambda _{s,\nu }\left( q\right) .$ 
}
\label{fig:fig5}
\end{figure}

From the calculated phonon dispersions in Fig. 5, we first notice that the
structure is stable. Secondly, we notice that a low frequency (1.8 THz) mode
exists near the point M. This mode is quadrupolar with the oxygen squares
rotated around the copper atoms. It can also be viewed as an in-plane
buckling mode and is strongly coupled to the in-plane $pd\pi $ bands. The
mode also involves displacements of the Ca atoms in the z-direction. It is
interesting to note that for the structure with flat planes, this
quadrupolar mode is unstable (at M, $\omega $ =3.7$i$ THz). This is
connected with the atypical feature near point A of the band structure in
Fig. 1. Here again we see that, for structural stability of this compound,
out-of-plane distortion is required.

\subsection{Linear electron-phonon interaction}

A large amount of theoretical work has been done to estimate the EPI in the
cuprates using the frozen-phonon approach \cite{Krak,Liecht}, the
rigid-muffin-tin \cite{RMT}, and tight-binding approximations \cite{Zhao}.
In addition, there are numerous attempts to deduce the strength of the EPI
from experiments \cite{Res}. So far, no general agreement is reached,
although the LDA frozen-phonon calculations \cite{Krak,Liecht} for selected,
high-symmetry phonons in YBa$_2$Cu$_3$O$_7$ and optimally doped La$_2$CuO$_4$
indicate that $\lambda _s\sim $0.5-1.5 . Compared herewith, our
linear-response method is more accurate because it allows calculation for
arbitrary phonon wave vectors ${\bf q}$. We have checked this method on EPIs
for a large number of classical superconductors \cite{EPI} and the results
indicate that, granted the LDA, we can obtain reliable estimates of $\lambda 
$ for the high-T$_c$ materials.

Due to the recent experimental evidence for $d_{x^2-y^2}$ symmetry of the
gap \cite{corr}, we have calculated the strength of the EPI in a general $L$%
-channel. We use a standard expression for the mode dependent coupling: 
\begin{eqnarray}
\lambda _{L,\nu }\left( {\bf q}\right) &=&\frac 1{\pi N_L}\frac{\gamma
_{L,\nu }\left( {\bf q}\right) }{\omega _\nu ^2\left( {\bf q}\right) }
\label{e1} \\
\ &\approx &\frac 2{\omega _\nu \left( {\bf q}\right) N_L}\sum_{{\bf k}%
}\,\,\delta \left( \epsilon \left( {\bf k}\right) \right) Y_L^{*}\left( {\bf %
\hat k}\right) \left| g_\nu \left( {\bf k,k+q}\right) \right| ^2  \nonumber
\\
&&\times Y_L\left( \widehat{{\bf k+q}}\right) \delta \left( \epsilon \left( 
{\bf k+q}\right) \right)  \nonumber
\end{eqnarray}
where $\sum_{{\bf k}}\equiv \int_{BZ}d^3k/BZV$ is the average over the
Brillouin zone, 
\begin{equation}
N_L\equiv \sum_{{\bf k}}\,\,\delta \left( \epsilon \left( {\bf k}\right)
\right) \left| Y_L\left( {\bf \hat k}\right) \right| ^2  \label{e2}
\end{equation}
is the electronic '$L$-density of states' per spin at the Fermi level $%
\left( \epsilon _F\equiv 0\right) ,$ $\gamma _{L,\nu }\left( {\bf q}\right) $
is the phonon line-width due to EPI in the $L$-channel, $\omega _\nu \left( 
{\bf q}\right) $ is the phonon energy, and we have used Rydberg units. For
the electronic energies, $\epsilon _j\left( {\bf k}\right) ,$ we have
dropped the band index $j$ because the FS of CaCuO$_2$ has only one sheet.
The approximation in the second line of (\ref{e1}) is, that the bands are
linear in the range $\epsilon _F\pm \omega _\nu \left( {\bf q}\right) .$
This approximation ought to be scrutinized in a future work. Finally, the
coupling constant $\lambda _L\equiv \sum_{v{\bf q}}\lambda _{L,\nu }\left( 
{\bf q}\right) $ is the sum over all phonon branches $\nu $ and the average
over the Brillouin zone.

Fig. 5 shows $\lambda _{s,\nu }\left( {\bf q}\right) $ for the different
phonon branches as a function of ${\bf q}$ along high-symmetry directions.
The EPI is seen to be particularly large for the buckling modes (the two
uppermost of the branches marked by circles). Remember that for flat planes,
these contributions to the EPI would vanish. It is furthermore seen that the
EPI is strongly enhanced along $\Gamma $X, and peaks near $\Gamma $X/3
reaching a maximum of about 2.5 . This is partly due to nesting
corresponding to {\em sliding} of the nearly flat (100)- or (010)-oriented
part of the FS on itself, or on the part parallel to it and displaced by $%
(0,\frac \pi {3a},0)$ or $(\frac \pi {3a},0,0),$ respectively. These
sliding-nestings cause $\sum_{{\bf k}}\delta \left( \epsilon \left( {\bf k}%
\right) \right) \delta \left( \epsilon \left( {\bf k+q}\right) \right)
\propto \lim_{\omega =0}\chi _0^{\prime \prime }\left( {\bf q,}\omega
\right) /\omega $ to be large near the following flat regions of ${\bf q}$%
-space: ${\bf q=}\left( q_x,0,q_z\right) ,$ ${\bf q=}\left( 0,q_y,q_z\right)
,{\bf q=}\left( q_x,\frac \pi {3a},q_z\right) $ and ${\bf q=}\left( \frac
\pi {3a},q_y,q_z\right) ,$ with $q_x$ and $q_y$ considerably smaller than $%
\frac \pi a$ and $q_z$ considerably smaller than $\frac \pi c.$ This kind of
nesting enhancement is well-known in the theory of HTSC \cite{nest}. A
second reason for the peaking of $\lambda _{s,\nu }\left( {\bf q}\right) $%
near $\Gamma $X/3 is a particular wave-vector dependence the EP coupling
constants $g_\nu \left( {\bf k,k+q}\right) $ for the buckling modes; this we
shall return to in the next section where we consider the tight-binding
model. From Fig. 5 we see that the EPI is small in other points of the
Brillouin Zone. Averaging over 15 ${\bf q}$-points in the $\Gamma $XM-plane
yields: $\lambda _s(q_z$=0$)=0.5,$ and averaging over the ZRA-plane gives: $%
\lambda _s(q_z$=$\frac \pi c)=0.2$. These values are respectively upper and
lower bounds for $\lambda _s$ averaged over all ${\bf q}$. We conclude that $%
\lambda _s\sim 0.4$\ is too small to account for the high-temperature
superconductivity in doped CaCuO$_2.$

\begin{figure}[t]
\centerline{\epsfysize=10cm \epsffile{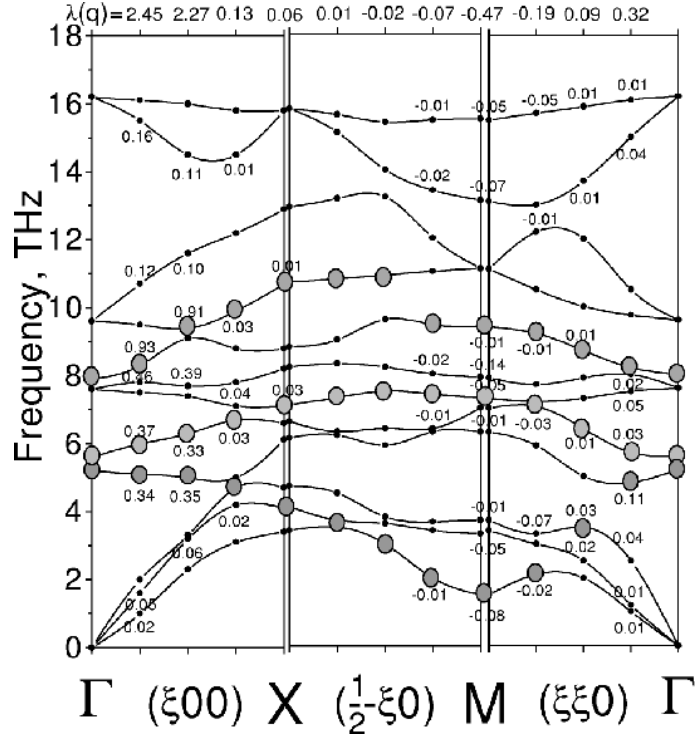}}
\caption{
Contributions to the electron--phonon
coupling in the $d_{x^2-y^2}$--channel from different phonon branches. See
also caption to Fig. 5.
}
\label{fig:fig6}
\end{figure}

Finally, we discuss our calculation of the $d$-channel EPIs. Using eqs. (\ref
{e1}-\ref{e2}) we have analyzed all symmetry-allowed channels of
superconductivity and came to the conclusion that, apart from the $s$%
-channel, relatively large positive values of $\lambda _L$ are only obtained
for pair-function symmetry $d_{x^2-y^2}$. Note that if $g$ were a constant,
independent of ${\bf k}$ and ${\bf q,}$ $\lambda _L$ would vanish for $L\neq
0.$ The $\lambda _{x^2-y^2,v}\left( {\bf q}\right) $are presented in Fig. 6
and may be compared with $\lambda _{s,\nu }\left( {\bf q}\right) $ in Fig.
5. Also for $\lambda _{x^2-y^2,v}\left( {\bf q}\right) $ there is a strong
enhancement near $\Gamma $X/3 and, in particular, for the buckling modes. It
is also seen that near the M-point, $\lambda _{x^2-y^2,v}\left( {\bf q}%
\right) $ vanishes for the buckling modes and is negative for all other
modes. Averaging over ${\bf q}$ in the $\Gamma $XM and ZRA planes give
respectively: $\lambda _{x^2-y^2}\left( q_z\rm{=0}\right) =0.4$ and $%
\lambda _{x^2-y^2}(q_z$=$\frac \pi c)=0.2.$ These values are {\em positive }%
and only{\em \ slightly smaller }than the corresponding values of $\lambda
_s $. The electron-phonon coupling is thus{\em \ highly anisotropic} in 24
per cent hole-doped CaCuO$_2$, and probably in all HTSCs near optimal doping.

\section{ANALYTICAL\ MODEL}

\subsection{Electronic 8-band Hamiltonian}

In the bottom part of Fig. 7 we specify an electronic 8-band orthogonal
tight-binding Hamiltonian for a single, dimpled CuO$_2$ plane, which we feel
is generic for CuO$_2$-HTSC's and contains the relevant degrees of freedom.
In the upper part, we synthesize the eight plane-bands. This Hamiltonian was
originally \cite{An94} derived from the LDA bands of YBa$_2$Cu$_3$O$_7$ by
integrating out the very-high-energy degrees of freedom, neglecting the
chains, and projecting onto one plane (or, for a bi-layer, to the even or
the odd linear-combination, or, for an infinite-layer, to a single $k_z$) 
\cite{An95}. We have recently \cite{Indra} deduced the parameter values of
this Hamiltonian for a number of further HTSC's and showed that (apart from
Fermi-level position) the only strongly material-dependent parameters are
the energy $\epsilon _s$ of the Cu$_{s-\left( 3z^2-1\right) }$ hybrid \cite
{An95}, which depends on the distance of apical oxygen from plane copper,
and the integral of hopping $t_{zd}$ from the Cu$_{x^2-y^2}$ orbital to an O$%
_z$ orbital, which is proportional to the buckling-angle $\delta .$ For
simplicity, we assume tetragonal symmetry in Fig.7 as well as in most of the
following. For a discussion of the inter-plane hopping, we refer to Ref. 
\cite{An95}.

\begin{figure}
\centerline{\psfig{file=fig7.epsi,height=14.5cm}}
\caption{
The 8-band Hamiltonian for a
single CuO$_2$ plane and synthesis of its band structure. The 1st (left)
column shows uncoupled $pd\sigma $ (O$_x$-Cu$_{x^2-y^2}$-O$_y),$ Cu$_s,$ and 
$pd\pi $ (Cu$_{xz}$-O$_z$ and Cu$_{yz}$-O$_z,$ stippled) bands. In the 2nd
(middle) column, the coupling $(t_{sp})$ of the Cu$_s$ orbital to the $%
pd\sigma $ band has been included. In the 3rd (right) column, also the
coupling $\left( t_{zd}\right) $ between $\sigma $- and $\pi $-bands has
been included. All energies are in eV.
}
\label{fig:fig7}
\end{figure}

Shown in the 2nd row and 1st column of Fig. 7, which we shall refer to as
Fig. 7(2,1), are the four $\sigma $-orbitals looked upon from above the
plane, $\left| y\right\rangle \equiv $ O$_y,$ $\left| d\right\rangle \equiv $
Cu$_{x^2-y^2},$ $\left| x\right\rangle \equiv $ O$_x,$ and $\left|
s\right\rangle \equiv $ Cu$_{s-\left( 3z^2-1\right) }.$ Note that O$_y$ and O%
$_x$ refer to orbitals on two different oxygens, O2 at $(1,0,\tan \delta
)a/2 $ and O3 at $(0,1,\tan \delta )a/2$. For some purposes, we shall also
distinguish the two oxygens by labels $a$ and $b.$ In Fig. 7(3,1), we see
the $\left| d\right\rangle $ orbital and two of the $\pi $-orbitals, $\left|
z\right\rangle \equiv $ O$_z$ and $\left| xz\right\rangle \equiv $ Cu$_{xz}$
from the edge of the plane. Approximate orbital energies (in eV and with
respect to the energy of the Cu$_{x^2-y^2}$ orbital) are given at the
relevant points of the band structure shown above, in Fig.7(1,1). Hence, $%
\epsilon _x$=$\epsilon _y\equiv \epsilon _p$ is 0.9 eV {\em below} $\epsilon
_d,$ and $\epsilon _s$ is 6.5 eV above. Moreover, $\epsilon _{za}$=$\epsilon
_{zb}\equiv \epsilon _z$ is 0.4 eV above $\epsilon _d,$ and $\epsilon _{xz}$=%
$\epsilon _{yz}$ is 1 eV below. If we now, as shown in Figs. 7 (2,1) and
(3,1) include the $pd\sigma $ hopping integrals $t_{xd}$=$t_{yd}\equiv
t_{pd} $=1.6 eV and the $pd\pi $ hopping integrals $t_{z,xz}$=$t_{z,yz}$=0.7
eV, we obtain the band structure shown in Fig. 7(1,1): The $\sigma $%
-orbitals give rise to a bonding, a non-bonding, and an anti-bonding O$_x$%
--Cu$_{x^2-y^2}$--O$_y$ band plus a Cu$_{s-\left( 3z^2-1\right) }$ level
(full lines), and the $\pi $-orbitals give rise to two decoupled pairs of
bonding anti-bonding bands which disperse in either the $x$ or $y$ direction
(stippled lines). The anti-bonding $pd\sigma $ band, which will develop into
the conduction band, has saddle-points at X (and Y) which are well above the
top of the $\pi $-bands, and which are {\em isotropic} in the sense, that
the absolute values of the band masses in the $x$ and $y$ directions are
equal. This means that the FS at half-filling is a square with corners at X
and Y. This is shown in Fig. 8(1,1).

In the 2nd columns of Figs. 7 and 8, the $\sigma $ and $\pi $ bands remain
decoupled, but we have introduced the Cu$_s$--O$_x$ and Cu$_s$--O$_y$
hoppings ($t_{sp}$=2.3 eV), as well as the tiny O$_z$--O$_z$ hopping (t$%
_{zz} $=0.06 eV). The latter is the only one reaching beyond nearest
neighbors and it merely lifts a degeneracy of the anti-bonding $\pi $-bands
at $\Gamma .$ We shall neglect it in the following. The {\em strong }%
coupling of the {\em remote }Cu$_{s-\left( 3z^2-1\right) }$ orbital to the $%
pd\sigma $ orbitals has the pronounced effect of depressing the conduction
band near the saddle-points at X and Y, and thereby increasing the mass
towards $\Gamma $ and decreasing it towards M $\left( \pi ,\pi \right) .$
This kind of saddle-point we refer to as {\em anisotropic. }As seen in Fig.
8(1,2), the FS passing through X and Y will therefore correspond to a finite
hole-doping and will bulge towards $\Gamma $. That the Cu$_{s-\left(
3z^2-1\right) }$ orbital, which has the azimuthal quantum number $m_z$=0,
may mix with the conduction band at X (and Y), but not along $\Gamma $M
follows from the symmetries of the corresponding two antibonding $pd\sigma $
wave functions \cite{An95}. The net downwards shift of the saddle-point
energy is the result of down-pushing by Cu$_s$ and weak up-pushing by Cu$%
_{3z^2-1},$ whose energy is below the saddle-point.

As a result of the hybridization with the Cu$_{s-\left( 3z^2-1\right) }$
orbital, the saddle-points of the anti-bonding $\sigma $-band straddle off
the top of the appropriate $\pi $-band, so that even a weak dimple or buckle
of the planes will introduce noticeable hybridization between the $\sigma $
and $\pi $ bands. This is seen in the 3rd column of Fig. 7 where we have
turned on the weak Cu$_{x^2-y^2}$--O$_z$ hoppings ($t_{zd}$=0.24 eV $\propto
\sin \delta $). These couplings become allowed when there is a finite angle $%
\delta $ ($\approx 7^o$ in YBa$_2$Cu$_3$O$_7,$ YBa$_2$Cu$_4$O$_8,$ and in
calculated CaCuO$_2)$ between the Cu-O bond and the plane of the
2D-translations. Since the hybridization between the antibonding $pd\sigma $
and Cu$_{xz}$--O$_z$ $pd\pi $ orbitals vanish at X (along the XM-line), the $%
\sigma $-$\pi $ hybridization may make the saddle-point {\em bifurcate} away
from X. This is seen in Fig. 8(1,3). At X (along XM), the $pd\sigma $-band
does hybridize with the more remote Cu$_{yz}$--O$_z$ $pd\pi $ band.

\subsection{Constant-energy contours (CECs)}

The orbitals in the Bloch representation and the 8-band Hamiltonian may be
found in eq.(2) of Ref. \cite{An94} and in eq.(1) of Ref. \cite{An95}. As
shown in the former reference, and given explicitly in its eqs.(12-16), the
constant-energy contours (CECs), $\epsilon _j\left( {\bf k}\right) =\epsilon
,$ have simple analytical expressions. These are given in the 3rd and 4th
rows of Fig. 8 in order of increasing couplings as in Fig. 7. In the present
paper we shall neglect the Cu$_{xz}$ and Cu$_{yz}$ orbitals and, hence,
consider the 6- rather than the 8-band Model (this corresponds to using
eq.(14) in Ref.\cite{An94}), because then the CECs are just hyperbolas in $%
\left( x,y\right) $-space, where 
\begin{equation}
x\equiv \frac 12\left( 1-\cos ak_x\right) \quad \rm{and\quad }y\equiv
\frac 12\left( 1-\cos ak_y\right) ,\quad  \label{e3}
\end{equation}
and they are shown in the 2nd row of Fig. 8. A CEC is specified by 3 numbers 
$d,$ $s,$ and $p,$ which are the values of the scattering functions defined
in the 4th row of Fig. 8. $d\left( \epsilon \right) $ specifies the $%
pd\sigma $ interaction, $s\left( \epsilon \right) $ the $sp\sigma $
interaction, and $p\left( \epsilon \right) $ the $\sigma \pi $ interaction.
Only $d\left( \epsilon \right) $ has a non-negligible (increasing)
energy-dependence in the relevant $\pm 0.1$ eV range around $\epsilon _F.$
For isotropic saddlepoints, $s\rightarrow \infty ,$ and for the cases of
interest, $s\sim 0.5-1.5.$ For a flat plane, $p=0.$ If $p=s^2/\left(
1+s\right) ^2,$ the saddlepoint is {\em extended,} that is, the dispersion
is quadratic towards M, but {\em quartic }towards $\Gamma .$ If $p$ exceeds
this value, the saddlepoint bifurcates to the positions given by $x=s\left(
p^{-1/2}-1\right) $ and $y=0.$ It should be noted that, although the CECs
near $\epsilon _F$ can be well reproduced by the 6-band model with an
energy-independent $p,$ because $\epsilon _{xz}$=$\epsilon _{yz}$ are
several eV below $\epsilon _F$ and 1.4 eV below $\epsilon _{za}$=$\epsilon
_{zb}$, the value of $p$ should be adjusted to give the correct CECs.

The structure of $\chi _0^{\prime \prime }\left( {\bf q,}\omega \right) $
obtained from the 6-band model as functions of doping, $s,$ and $p$ is
discussed in Ref. \cite{Jeps}.

The formalism given above is for tetragonal symmetry and apparently only
applies to a plane dimpled like in YBa$_2$Cu$_3$O$_7,$ but not to a plane
buckled as in Fig. 2. However, since in the 6-band model each O$_z$-orbital
couples only to the Cu$_{x^2-y^2}$-orbital, we can freely define the sign of
the O$_z$ orbital at $(1,0,\tan \delta )a/2,$ $\left| za\right\rangle ,$ to
be the opposite of that at $\left( 0,1,-\tan \delta \right) a/2$, $\left|
zb\right\rangle ,$ and then the formalism holds also for static buckling.
Nevertheless, for the sake of preciseness in the wording, we shall assume
static dimpling in the following. The values of the electronic parameters
for various materials may be found in Ref.\cite{Indra}.

\begin{figure}
\centerline{\psfig{file=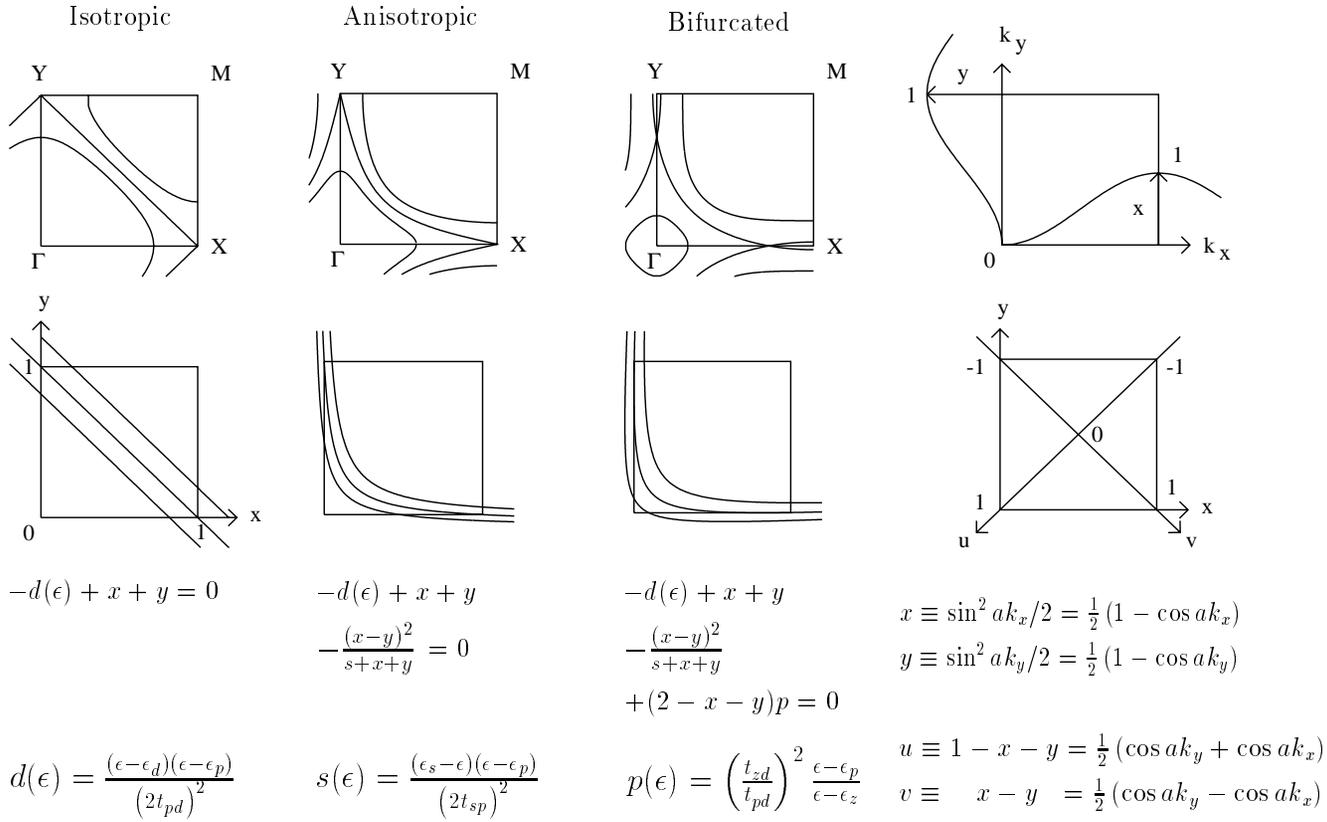,height=17.5cm}}
\caption{
Caption is given at the next page.
}
\label{fig:fig8}
\end{figure}

In $\left( x,y\right) $-space the average over the Brillouin zone becomes: 
\begin{equation}
\sum_{{\bf k}}\equiv \left( \frac {a}{2\pi }\right) ^2\int_{-\pi /a}^{\pi
/a}\int_{-\pi /a}^{\pi /a}dk_x\,dk_y=\left( \frac {1}{\pi }\right)
^2\int_0^1\int_0^1\frac{dxdy}{\sqrt{xy\left( 1-x\right) \left( 1-y\right) }}%
.  \label{e4}
\end{equation}

CAPTION to Fig. 8. The top row of the first three columns shows CECs in $%
\left( k_x,k_y\right) $-space of the 8-band Hamiltonian corresponding to the
three columns in Fig. 7. The CECs chosen in the figures are antibonding $%
pd\sigma $-like and are, specifically, the one passing through the
saddle-point $\left( \epsilon \equiv \epsilon _{\rm{saddle}}\right) $ and
two neighboring ones $\left( \epsilon \equiv \epsilon _{\rm{saddle}}\pm
\Delta \epsilon \right) $. The 4th column illustrates the transformation
from $\left( k_x,k_y\right) $-space to $\left( x,y\right) $-, or $\left(
u,v\right) $-, space. The 2nd row shows the CECs in $\left( x,y\right) $%
-space where they are hyperbolas. The 3rd row gives the expressions for the
CECs and the $\left( k_x,k_y\right) $-$\left( x,y\right) $ transformation.
The 4th row gives the expressions for the scattering functions, and the $%
\left( u,v\right) $-$\left( x,y\right) $transformations. For simplicity, the
CECs in the 3rd column are for the 6-band Hamiltonian where, of the $\pi $%
-orbitals, only the two O$_z$-orbitals and not the Cu$_{zx}$- and Cu$_{yz}$%
-orbitals are included.\bigskip\ 

\subsection{Coupling of electrons to the buckling mode}

The displacement pattern of the buckling mode is: 
\begin{equation}
\left\{ \frac{\partial {\bf R}_{\rm{O2}}}{\partial Q_{{\bf q}}},\frac{%
\,\partial {\bf R}_{\rm{O3}}}{\partial Q_{{\bf q}}}\right\} =\frac{{\bf %
\hat z}}{\sqrt{2M\omega }}\frac 1{\sqrt{2}}\left\{ \exp \left( -i{\bf q\cdot
R}_{\rm{O2}}\right) ,\,-\exp \left( -i{\bf q\cdot R}_{\rm{O3}}\right)
\right\}  \label{e5}
\end{equation}
where O2 and O3 are the oxygens at respectively $(1,0,\tan \delta )a/2$ and $%
\left( 0,1,\tan \delta \right) a/2$ and ${\bf q}$ is 2-dimensional.

For ${\bf q=0,}$ this is the mode for{\bf \ }which the dimple of O2 is
increased and that of O3 is decreased, or vice versa. For YBa$_2$Cu$_3$O$_7,$
this is the 330 cm$^{-1}$ out-of-phase mode. Had $\delta $ been zero at
equilibrium, oxygen would have been dimpling upwards in the --Cu--O2--Cu--
rows and downwards in the --Cu--O3--Cu-- rows, that is what we have called
buckling.

${\bf q=}\left( \frac \pi a,0\right) :$ Had $\delta $ been zero at
equilibrium, this mode would have been buckling in the $x$-direction and
dimpling in the $y$-direction.

${\bf q=}\left( \frac \pi a,\frac \pi a\right) :$ Had $\delta $ been zero at
equilibrium, both the --Cu--O2--Cu-- and the --Cu--O3--Cu--rows
would be buckling, and such that the O2--O3--O2--rows in the {\it x}=%
{\it y} direction are in phase. This is the displacement pattern of the
low-temperature orthorhombic (LTO) phase in La$_2$CuO$_4.$

For this mode, the electron-phonon matrix element is: 
\[
g\left( {\bf k,\,k}^{\prime }\equiv {\bf k+q}\right) =
\]
\[
=\frac 1{\sqrt{2M\omega }}\left\langle {\bf k}\right| \sum_{{\bf R}}\frac 1{%
\sqrt{2}}\left( \frac{\partial H}{\partial z_{\rm{O2}}}\exp \left( -i{\bf %
q\cdot R}_{\rm{O2}}\right) -\frac{\partial H}{\partial z_{\rm{O3}}}\exp
\left( -i{\bf q\cdot R}_{\rm{O3}}\right) \right) \left| {\bf k}^{\prime
}\right\rangle 
\]
\[
=\frac{\partial t_{zd}/\partial z_{\rm{O}}}{\sqrt{M\omega }}\left[ 
\begin{array}{c}
c_d\,\left( {\bf k}\right) \left( \cos \frac a2k_x\,\,\left| c_a\left( {\bf k%
}^{\prime }\right) \right| -\cos \frac a2k_y\,\left| \,c_b\left( {\bf k}%
^{\prime }\right) \right| \right) + \\ 
{\rm sgn}\left\{ c_a\left( {\bf k}^{\prime }\right) c_a\left( {\bf k}%
\right) \right\} \left( \left| c_a\left( {\bf k}\right) \right| \cos \frac
a2k_x^{\prime }-\left| c_b\left( {\bf k}\right) \right| \cos \frac
a2k_y^{\prime }\right) \,c_d\left( {\bf k}^{\prime }\right) 
\end{array}
\right] 
\]
where, for simplicity, we have only included the phonon-induced modulation
of the hopping integral $t_{zd}.$ The modulation of $\epsilon _z$ due to
screening is considered in Ref. \cite{Indra}. The eigenvector-components, $%
c_d\left( {\bf k}\right) ,\,c_a\left( {\bf k}\right) ,$ and $c_b\left( {\bf k%
}\right) ,$ for the Bloch orbitals $\left| d,{\bf k}\right\rangle ,\,\left|
za,{\bf k}\right\rangle $ and $\left| zb,{\bf k}\right\rangle $ are real due
to our choice of phases for the Bloch functions, and they are most easily
obtained by differentiation of the CEC-equation with respect to the orbital
energies, $\epsilon _d,$ $\epsilon _{za},\,$and $\epsilon _{zb}.$ According
to first order perturbation theory, this yields $c_d\left( {\bf k}\right)
^2=\partial \epsilon \left( {\bf k}\right) /\partial \epsilon _d,$ and so
on. Simple considerations finally yield the relative signs. As a result, we
obtain by using the 6-band model: 
\begin{eqnarray*}
g\left( {\bf k,k}^{\prime }\right)  &=&2t_{zd}\,\frac{\epsilon _F-\epsilon _p%
}{\left( \epsilon _F-\epsilon _z\right) \left( \epsilon _F-\frac{\epsilon
_p+\epsilon _d}2\right) }\,\frac{\partial t_{zd}/\partial z_{\rm{O}}}{%
\sqrt{M\omega }}\;\times  \\
&&\left( \left| \cos \frac a2k_x\cos \frac a2k_x^{\prime }\right| -\left|
\cos \frac a2k_y\cos \frac a2k_y^{\prime }\right| \right) 
\end{eqnarray*}
\begin{equation}
=c\,\left( \sqrt{\left( 1-x\right) \left( 1-x^{\prime }\right) }-\sqrt{%
\left( 1-y\right) \left( 1-y^{\prime }\right) }\right) .  \label{e6}
\end{equation}
where $c$ is the constant in front of the parenthesis on the first line.

In order to get a feeling for the behavior of the pair-interaction, $%
V\propto -g^2,$ let us first take ${\bf q=0,}$ i.e. ${\bf k}^{\prime }={\bf %
k.}$ Eq.(\ref{e6}) then yields: $g=c\left( y-x\right) $ which shows that $%
g^2 $ attains its maximum $\left( c^2\right) $ for ${\bf q}$ small and ${\bf %
k}$ at those points of the FS where the density of states (distance between
neighboring CECs) is largest and where the superconducting gap is observed
to be largest. That is, near the saddlepoints.

Next, let us take ${\bf k}^{\prime }$ at the X-point, $\left[ x^{\prime
},y^{\prime }\right] =\left[ 1,0\right] .$ Then, $g=c\sqrt{1-y}.$ We thus
see that not only does $g^2$ take its {\em maximal} value for ${\bf k=k}%
^{\prime }$ at X, but it {\em vanishes} for ${\bf k}$ at X and ${\bf k}%
^{\prime }$ at Y. This is exactly what is needed for the electron-phonon
interaction to support $d_{x^2-y^2}$-pairing.

Finally, let ${\bf k}$ and ${\bf k}^{\prime }$ to be symmetric around the $%
\left( 1,1\right) $ or $\left( 1,-1\right) $ line, {\it i.e.} $\left(
k_x^{\prime },k_y^{\prime }\right) =$ $\left( \pm k_y,\pm k_x\right) ,$ in
which case ${\bf q}$ is in the $\left( 1,\mp 1\right) $ direction. In this
case, $g=V=0.$ In particular, this means that $V$ vanishes for ${\bf q=}%
\left( \frac \pi a,\frac \pi a\right) $ where the Coulomb repulsion is
maximal.

Having realized that the buckling mode has the right kind of $\left( {\bf k,k%
}^{\prime }\right) $-dependence to support $d_{x^2-y^2}$-wave pairing, let
us calculate $\lambda _{x^2-y^2}$ for the buckling mode and see whether it
is really positive. For this purpose we note that 
\begin{eqnarray*}
&&g^2\left( {\bf k,k}^{\prime }\right) =c^2\times \\
&&\left[ \left( 1-x\right) \left( 1-x^{\prime }\right) +\left( 1-y\right)
\left( 1-y^{\prime }\right) -2\sqrt{\left( 1-x\right) \left( 1-y\right)
\left( 1-x^{\prime }\right) \left( 1-y^{\prime }\right) }\right]
\end{eqnarray*}
is factorized so that we can easily calculate $\lambda _L\equiv \sum_{{\bf q}%
}\lambda _L\left( {\bf q}\right) $ as a double-integral, not over ${\bf k}$
and ${\bf q}$ as in (\ref{e1}), but over ${\bf k}$ and ${\bf k}^{\prime }.$
Dropping the ${\bf q}$-dependence of the buckling-mode frequency, we get
using (\ref{e4}): 
\begin{eqnarray*}
&&\lambda _s=\frac{4c^2}{\pi ^2\omega }\times \\
&&\frac{\left[ \int_0^1\int_0^1\left( 1-x\right) \delta \left( \epsilon
\right) \frac{dxdy}{\sqrt{xy\left( 1-x\right) \left( 1-y\right) }}\right]
^2-\left[ \int_0^1\int_0^1\delta \left( \epsilon \right) \frac{dxdy}{\sqrt{xy%
}}\right] ^2}{\int_0^1\int_0^1\delta \left( \epsilon \right) \frac{dxdy}{%
\sqrt{xy\left( 1-x\right) \left( 1-y\right) }}}
\end{eqnarray*}
where $\delta \left( \epsilon \right) \equiv \delta \left( \epsilon \left(
x,y\right) -\epsilon _F\right) .$ Using $Y_{x^2-y^2}\left( {\bf k}\right)
\propto x-y$ we get: 
\begin{eqnarray*}
&&\lambda _d=\frac{4c^2}{\pi ^2\omega }\times \\
&&\frac{\left[ \int_0^1\int_0^1\left( 1-x\right) \left( x-y\right) \delta
\left( \epsilon \right) \frac{dxdy}{\sqrt{xy\left( 1-x\right) \left(
1-y\right) }}\right] ^2-\left[ \int_0^1\int_0^1\left( x-y\right) \delta
\left( \epsilon \right) \frac{dxdy}{\sqrt{xy}}\right] ^2}{%
\int_0^1\int_0^1\left( x-y\right) ^2\delta \left( \epsilon \right) \frac{dxdy%
}{\sqrt{xy\left( 1-x\right) \left( 1-y\right) }}}.
\end{eqnarray*}
Now, the last term of $\lambda _d$ vanishes and we may symmetrize the first
term as follows: $\left( 1-x\right) \left( x-y\right) \rightarrow \frac
12\left[ \left( 1-x\right) \left( x-y\right) +\left( 1-y\right) \left(
y-x\right) \right] =-\frac 12\left( x-y\right) ^2$. As a result: 
\begin{equation}
\lambda _d=\frac{c^2}{\pi ^2\omega }\int_0^1\int_0^1\left( x-y\right)
^2\delta \left( \epsilon \left( x,y\right) \right) \frac{dxdy}{\sqrt{%
xy\left( 1-x\right) \left( 1-y\right) }}=\frac{c^2}\omega \,N_d  \label{e7}
\end{equation}
which is {\em positive}, and easy to evaluate.

It is obvious from these equations, that if the FS had weight merely at X
and Y, that is, if $\epsilon _F$ were at a non-bifurcated saddlepoint, then 
\[
\lambda _s=\lambda _d\;\left( \rightarrow \infty \right) . 
\]
For $\epsilon _F$ near an extended saddlepoint, $\lambda _d$ is particularly
large.

The integral (\ref{e7}), as well as the one for the normal density of
states, $N\equiv N_s,$ may be expressed analytically in terms of elliptic
integrals if we use a band structure with $p=0,$ i.e. the 4-band model. The
results for $s=100$ and $s=0.6$ are shown graphically in Fig. 9 as a
functions of the doping. We see that for doping to the van Hove singularity, 
$N_d=N,$ as was stated above.

\begin{figure}
\centerline{\psfig{file=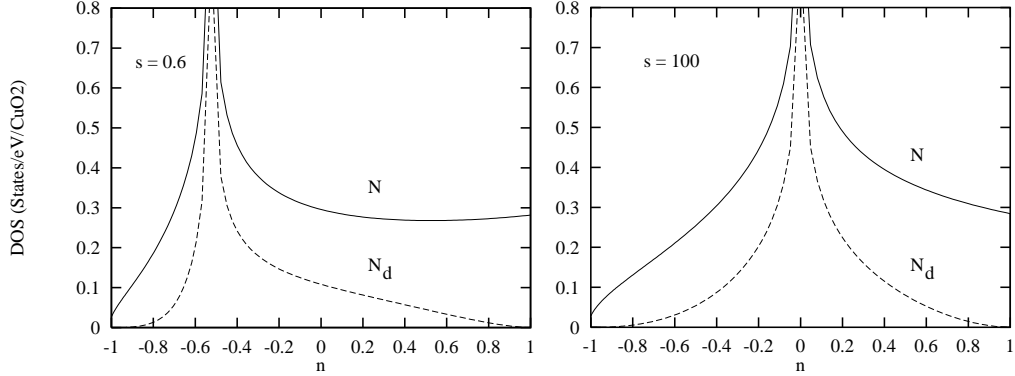,height=5.0cm}}
\caption{
Density of states and $
d_{x^2-y^2}$ weighted density of states as defined in eq. 2 for the 4-band
model. $n$ is the number of electrons counted from the half-full band.
}
\label{fig:fig9}
\end{figure}

\section{CONCLUSIONS}

We conclude that the linear electron-phonon interaction for buckled planes
may support, but is hardly sufficient to cause, high-temperature
superconductivity based on $d_{x^2-y^2}$-pairing. The most important mode
seems to be the buckling mode, because it modulates the saddlepoints of the
energy bands where the density of states is high and where the
superconducting gap is observed to be maximum. Moreover, the buckling mode
does not interact with the electrons for ${\bf q}$ near $\left( \frac \pi
a,\frac \pi a\right) $ where the Coulomb repulsion is maximum.

Considering the facts that the linear EPI for the buckling mode is
proportional to the small static (on the scale of phonon frequencies)
buckling, and that this mode is calculated to be highly anharmonic, it seems
imminent to investigate the role of anharmonicity. Similarly, the interplay
between the electron-phonon interaction and the Coulomb repulsion should be
studied. For these purposes the analytical 6-band models seem realistic and
tractable.

\section*{ACKNOWLEDGMENTS}

The authors are indebted to M. Kuli\'c, E. G. Maksimov, I. I. Mazin, and R.
Zeyer for many helpful discussions.

\end{document}